# Strain-Tuned Optical Properties of a Two-Dimensional Hexagonal Lattice: Exploiting Saddle Degrees of Freedom and Saddle Filtering Effects

Phusit Nualpijit [a,b], Bumned Soodchomshom[a,*]

[a] Department of physics, Faculty of science, Kasetsart University, Bangkok 10900, Thailand
[b] School of *Integrated* Science (SIS), Kasetsart University, Bangkok 10900, Thailand
*Corresponding Author's E-mail: fscibns@ku.ac.th, bumned@hotmail.com

**ABSTRACT**: The deformation of hexagonal lattices has attracted considerable attention due to its promising applications in straintronics. This study employs the tight-binding model to investigate the anisotropic spectrum, where electronic transport can be manipulated by the degree of deformation. The longitudinal conductivities, light transmittance, and absorbance are analyzed, revealing enhancement along one direction and suppression along the other. The findings indicate that the direction and magnitude of strain can be determined by measuring transmittance and absorbance, showing significant deviations from the unstrained condition. Furthermore, a strong absorbance is observed due to the interband transition of electrons near the M-point saddles, linked to van Hove singularities when $\omega = 2|t'|$ and $\omega = 2|2t - t'|$, where t and t' are nearest and next-nearest hoping energy, respectively. The unexpected characteristics of saddle polarization— analogous to valley polarization at K and K'—become particularly prominent when strain affects the selection of M-point saddle. Notably, the demonstration indicates that a highly efficient M-point saddle filtering effect takes place, induced by linearly polarized light. This model paves the way for exploring the optical properties of anisotropic hexagonal lattices, such as black phosphorus and borophene oxide. These results also open a pathway to **strain-programmable optoelectronic devices**, such as polarization-selective photodetectors, tunable absorbers, and ultrathin optical filters.

## I. INTRODUCTION.

Monolayer materials composed of a single element and arranged in a hexagonal lattice have been extensively studied, with particular attention to group IIIA and IVA elements - such as graphene, silicene, borophene, collectively known as "Xenes" [1-3]. Typically, the electronic spectrum exhibits a well-defined linear dispersion near two valleys, $K$ and $K'$, known as Dirac cones. Basically, the prototype material, graphene, exhibits a gapless semimetal, protected by time-reversal and spatial inversion symmetries. These symmetries can be broken by spin doping [4], bicirularly polarized light [5, 6], and the orbital magnetic moment induced by a gate voltage in multilayer graphene [7, 8], which contribute to the formation of a tunable band gap. Strain engineering facilitates the modification of Slater-Koster parameters, leading to anisotropic spectral characteristics and transport properties [9]. However, under sufficiently high pressure, the two cones merge, producing a modified spectrum that retains linearity in one direction but quadratic in the other [10-12].

The optical transport properties exhibit anisotropy when the polarized light illuminates the surface of a uniformly deformed lattice. The transmittance and absorbance can be modulated by the applied tension and polarization direction of the incident light, enabling the determination of both the magnitude and direction of the tension [13]. A small lattice deformation leads to a slight deviation in transmittance and absorbance, with transmittance varying around 97.7% across a broad frequency range [14]. In contrast, under strong lattice deformation, transmittance can significantly deviate- from nearly 0% to almost 100% - depending on the polarization direction within the microwave and far-infrared regimes [15-17]. This phenomenon arises from enhancement in one direction and suppression in the other [12].

Recently, Natalia Cortes et al. investigated the impact of biaxial tensile strain on phosphorus-doped graphene monolayer [18]. Their findings reveal the emergence of a magnetic state, associated with the out-of-plane P–C bond, characterized by a $sp^3$-like hybrid orbital. When the mechanical force is applied, a transition from magnetic state into nonmagnetic state is observable. This phenomenon occurs as increasing strain, the $sp^2$ hybrid orbital is recovered, leading to the formation of a flat hexagonal lattice. These results align with theoretical studies on spin-spin interactions of substitutional atoms on hexagonal lattices, as detailed in Ref. [19]. Additionally, their analysis highlights that the density of states express the dominant peak near the Fermi level, resulting from the anisotropic spectrum. This suggests that the strength of spin-spin interactions can be modulated through strain parameters.

In this paper, we propose the model of uniaxial strain on hexagonal lattice based on the tight-binding model. Mechanical pressure is applied along the y-axis, resulting in variation of the hopping parameter $t'$, as illustrated in Fig.1. It is important to note that the variation of the Slater-Koster parameters of two $p_z$ orbitals are related to the interorbital distance $l$ by $V_{pp\pi}(l) = t\exp[-3.37(l/a-1)]$, where "a" represents the lattice constant and $t$ is the hopping parameter of the system in equilibrium [9]. Specifically, when $t'/t > 1$, compression occurs along the armchair orientation. Conversely, when $t'/t < 1$, tension is directed along the armchair direction. Additionally, the optical properties are examined through both analytical and numerical methods, revealing anisotropic behavior. The findings highlight the benefits of anisotropic materials in optical applications. Moreover, the unexpected characteristics of saddle polarization—analogous to the valley polarization at K and K'—become especially pronounced when strain influences the selection of M-point saddles. The demonstration states that a nearly perfect M-point saddle filtering effect occurs at $M_3$, driven by linearly polarized light. This novel of saddle polarization exhibits a contrasting behavior to valley polarization, which is governed by circularly polarized light, inducing an energy gap that varies with light intensity at K and K', as described by Floquet theory [20-23]. The model can be broadly applied to other materials, such as black phosphorus [24, 25], and borophene oxide [26, 27].

## II. MODEL

In this section, we examine the behavior of electrons within a hexagonal lattice, employing the tight-binding model, constrained to interactions with nearest-neighbor atoms. The generalized Hamiltonian is formulated as follows [1, 28]

$$H_0 = h_1\sigma_1 + h_2\sigma_2 \tag{1}$$

In this context, the relevant matrices are Pauli matrices $\sigma_i$ and the corresponding coefficients $h_1$, expressed in the momentum representation $\mathbf{q} = (q_1, q_2)$, associated with these matrices for the system are defined as

$$h_1 = -\sum_{i=1,2,3} t_i \cos(\mathbf{q}\cdot\mathbf{a}_i) \tag{2}$$

$$h_2 = -\sum_{i=1,2,3} t_i \sin(\mathbf{q}\cdot\mathbf{a}_i) \tag{3}$$

The nearest-neighbor vectors are given by $\mathbf{a}_1 = a(0,-1)$, $\mathbf{a}_{2,3} = a/2(\pm\sqrt{3},1)$, Here, $t_i$ represents the hopping parameters for each direction, while $a$ represents the lattice constant, as illustrated in Fig.1. The eigenvalues and eigenvectors of the Hamiltonian are expressed as follows

$$E_k = \pm\sqrt{h_1^2 + h_2^2} \tag{4}$$

$$|\psi_\pm\rangle = \frac{1}{\sqrt{2}}\begin{pmatrix} \pm 1 \\ \dfrac{h_1 + ih_2}{E_k} \end{pmatrix} \tag{5}$$

In case of free-standing hexagonal lattice, uniform hopping parameters yield a linear dispersion relation characterized by two Dirac nodes within each Brillouin zone [1]. Conversely, deformation induces variations in the hopping parameters across different directions. Breaking isotropy by increasing the hopping parameter along $t_1$, while keeping $t_2 = t_3 = t$, causes the two Dirac nodes to approach one another and eventually merge when $t_1 = 2t$ [10].

The low-energy Hamiltonian is obtained by expanding Eq. 1 around the vicinity of Dirac points $\mathbf{q} = \mathbf{K}_\pm + \mathbf{k}$ ($\mathbf{K}_D \ll \mathbf{k}$), whose positions in momentum space are altered due to deformation. In this consideration, the hopping parameters $t_2 = t_3 = t$ are kept constant, while another $t_1 \equiv t'$ is adjustable. Therefore, the nonequivalent Dirac points are positioned at

$$K_\pm = \left(\pm\frac{2}{\sqrt{3}}\cos^{-1}\left(-\frac{t'}{2t}\right), 0\right) \tag{6}$$

Therefore, the approximation when $t' < 2t$ reads [9]

$$h_1 \approx \eta\frac{t}{2}\sqrt{12 - \frac{3t'^2}{t^2}}k_1 \tag{7}$$

$$h_2 = \frac{3t'}{2}k_2 \tag{8}$$

and when $t' \geq 2t$ reads [11, 29]

$$h_1 = \frac{3t}{4}k_1^2 + \Delta \tag{9}$$

$$h_2 = (t + t')k_2 \tag{10}$$

where $\eta = \pm 1$ refers to the valley index, and the energy gap parameter $\Delta \equiv t' - 2t$ provides the energy band gap $2\Delta$ when $t'$ exceeds $2t$ [9]. The shifting of two Dirac points as the hopping parameter $t'$ varies are illustrated in Fig.2. For $t'/t < 1$, the two Dirac points deviate from the isotropic case, as in illustrated Fig. 2b. When $1 < t'/t < 2$, the nodal points approach one another, as depicted in Fig. 2d, and eventually merging at $t'/t = 2$, as depicted in Fig. 2e. When $t'/t > 2$, the gap begins to open at the same position in momentum space as observed in case of $t'/t = 2$.

## III. DENSITY OF STATES (DOS)

The density of states (DOS) for two dimensional lattice reads [30]

$$\rho(E) = \int_{BZ} \frac{d^2\mathbf{k}}{4\pi^2}\delta(E - E_k) \tag{11}$$

The integral is performed over the entire Brillouin zone (area within red diamond), as defined in Fig.2c. Note that the area of Brillouin zone remains unchanged regardless of changes in $t'$. The DOS is evaluated numerically in various conditions of $t'$, as illustrated in Fig.3a. This indicates that modifying the lattice deformation can effectively control the increase or decrease in the number of states that are available in the system. The van Hove singularity is generated from the saddle point, which are $M$ points. There are two conditions where the singularities are located : $E = |t'|$ from $M_1$, $M_2$ and $E = |2t - t'|$ from $M_3$, as defined in Fig.2c. When $t' \geq 2t$, $M_3$ point merges with $K$ and $K'$ to form a single node. Consequently, only one singularity is observed.

Here, we examine the analytical formulation of the DOS for $t'/t < 2$. The new variables $u \equiv (t/2)\sqrt{|2 - 3t'^2/(t^2)|}$ and $v = 3t'/2$ are introduced to transform the integral from

$(k_1, k_2) \rightarrow (h_1, h_2)$, from the Eq.7-8, using the Jacobian determinant $\mathbb{J}_1 = 1/|uv|$ appropriate for the polar coordinates. The result is ultimately obtained as

$$\rho(E) = 4\mathbb{J}_1\frac{|E|}{2\pi} \tag{12}$$

The prefactor $4$ is taken into account for spin and valley degeneracy. For the case of uniform hopping parameter, the Eq.12 reduces to DOS for graphene $\rho = 2|E|/(\pi v_F^2)$, where $v_F \equiv 3at/2$ is the Fermi velocity [1, 4]. Fig. 3b illustrates the scaling function of the DOS (Eq.12) for $t' < 2t$, under different conditions of $t'$. The graph demonstrates that the minimum number of states available for electrons occupied at each energy level is attained under the condition of $t' = \sqrt{2}t$.

When $t'/t > 2$, the new variables $u \equiv 3t/4$ and $\tilde{v} = (t + t')$ are defined, including the Jacobian determinant $\mathbb{J}_2 = 1/|2u\tilde{v}k_1|$. The result is then expressed in the form [11]

$$\rho(E) = \frac{2E}{8\pi^2 v\sqrt{u}}\int_0^{\pi/2}d\theta\frac{1}{\sqrt{E\cos\theta - \Delta}} \tag{13}$$

The integral cannot be implemented analytically because of the elliptic integral of the first kind. When $t' = 2t$, the energy gap $\Delta$ reduces to zero which the elliptic integral can be approximated by

$$\int_0^{\pi/2}d\theta\frac{1}{\sqrt{\cos\theta}} \sim 2.62 \tag{14}$$

This results in $\rho(E) \approx 0.32\sqrt{E}$ for $t' = 2t$ [31], which changes from linearity to square root behavior and is eventually disrupted by the energy gap when $t' > 2t$. There are no electrons occupying the states where the energy lies within the energy gap.

## IV. OPTICAL CONDUCTIVITIES

Under the framework of linear response theory, the optical conductivity can be calculated by Kubo formula [28, 32].

$$\sigma_{\mu\nu}(\omega) = -\frac{i}{h}\sum_{\lambda,\lambda'=\pm 1}\lim_{\alpha\to 0}\int_{BZ}\frac{d^2\mathbf{k}}{(2\pi)^2}\frac{f(E_{\lambda'}) - f(E_\lambda)}{E_\lambda - E_{\lambda'}}$$
$$\times\frac{\langle\psi_\lambda|j_\mu|\psi_{\lambda'}\rangle\langle\psi_{\lambda'}|j_\nu|\psi_\lambda\rangle}{E_\lambda - E_{\lambda'} + \omega + i\alpha} \tag{15}$$

The current operator is defined by $j_\mu = e \, \partial H / \partial k_\mu$, where $\mu = x, y$, operating on eigenstates $|\psi_\lambda\rangle$ with band index $\lambda = \pm 1$ and $E_\lambda$ is the eigenenergy. In this scenario, the Fermi level is positioned between the valence band maximum and the conduction band minimum, the Fermi-Dirac distribution function $f(E_l) = 1 / \left[ 1 + \exp\{\beta(E_l - E_F)\} \right]$ with the inverse thermal energy $\beta = 1 / (k_B T)$ is considered.

The longitudinal conductivities, calculated from full band Hamiltonian, are plotted under various conditions of $t'$ for both the real part $\sigma_{\mu\mu}{}'(\omega)$ and imaginary part $\sigma_{\mu\mu}{}''(\omega)$, as illustrated in Fig.4. These plots are normalized by the constant universal conductivity of graphene $\sigma_0 = \pi e^2 / (2h)$, calculated from $t' = t$ in dc-limit and under the low temperature regime $T \to 0$. The broken isotropy introduces two singularities in $\sigma_{yy}{}'(\omega)$: one originating from $M_1$ point with the light energy $\omega / t = 2|t'|$ and the other from $M_3$ point with the light energy $\omega / t = 2|2t - t'|$, under the conditions $t' < 2t$ and $t' \neq t$. In contrast, $\sigma_{xx}(\omega)$ exhibits a single singularity located at $\omega / t = 2|t'|$. Fig.4a-4b reveal opposing behaviors, with conductivity being amplified in one direction while diminished in the other [12]. When $t' = 2t$, the longitudinal conductivities in the low-energy limit deviate from their constant behavior, transitioning into square root behavior in $\sigma_{xx}{}'(\omega)$ and inverse square root behavior in $\sigma_{yy}{}'(\omega)$, respectively [15]. For $t' > 2t$, the electron transport is similar to that of $t' = 2t$, with the threshold energy matching the energy gap. Conversely, the transverse conductivity cannot be observed.

Here we consider the analytical expression of the longitudinal conductivities for $t'/t < 2$, the inter band matrix elements read

$$\langle \psi_\pm | j_x | \psi_\mp \rangle \langle \psi_\mp | j_x | \psi_\pm \rangle = \frac{3(4t^2 - t'^2) h_2{}^2}{4(h_1{}^2 + h_2{}^2)} \qquad (16)$$

$$\langle \psi_\pm | j_y | \psi_\mp \rangle \langle \psi_\mp | j_y | \psi_\pm \rangle = \frac{9t'^2 h_1{}^2}{4(h_1{}^2 + h_2{}^2)} \qquad (17)$$

The coordinates transformation to the polar coordinates is necessary to evaluate the conductivity in low-energy limit, which changes from $(k_1, k_2) \to (h_1, h_2)$. This gives rise to the Jacobian determinant $\mathbb{J}_1 = 1 / |uv|$ as same as in Eq.12. Therefore, the real part of the longitudinal conductivities per spin and valley projections, as a function of strain parameter $t_1$, can be expressed as

$$\sigma_{xx}{}'(\omega) = \mathbb{J} \frac{3(4t^2 - t'^2)}{4} \left( \frac{\sigma_0}{4} \right) \qquad (18)$$

$$\sigma_{yy}{}'(\omega) = \mathbb{J} \frac{9t'^2}{4} \left( \frac{\sigma_0}{4} \right) \qquad (19)$$

Conveniently, the imaginary components of the longitudinal conductivities $\sigma_{xx}{}''(\omega)$ and $\sigma_{yy}{}''(\omega)$ approach zero, as illustrated in Fig.4c-4d. It is important to note that the product of $\sigma_{xx}\sigma_{yy}$ yields square of the universal conductivity of graphene $\sigma_0{}^2$ [12, 15, 33], considering the contributions from spin and valley projections.

When $t' \geq 2t$, the inter band matrix elements read

$$\langle \psi_\pm | j_x | \psi_\mp \rangle \langle \psi_\mp | j_x | \psi_\pm \rangle = \frac{3t h_2 (h_1 - \Delta)}{h_1{}^2 + h_2{}^2} \qquad (20)$$

$$\langle \psi_\pm | j_y | \psi_\mp \rangle \langle \psi_\mp | j_y | \psi_\pm \rangle = \frac{h_1{}^2 (t + t')^2}{h_1{}^2 + h_2{}^2} \qquad (21)$$

The Jacobian determinant $\mathbb{J}_2 = 1 / |2u\tilde{v}k_1|$ is the same as in Eq.13. The longitudinal conductivity per spin projections, as a function of strain parameter $t_1$, can be expressed as

$$\sigma_{xx}(\omega) = \frac{i}{2\pi^2} \left( \frac{\sqrt{3t}}{(t + t')} \int d\theta d\varepsilon \, \frac{1}{2\varepsilon + \omega + i\alpha} \right.$$
$$\left. \times \frac{\varepsilon^{5/2} \sqrt{\cos\theta} \sin^2\theta}{\varepsilon^2 + \Delta^2 - 2\Delta\varepsilon\cos\theta} \right) \sigma_0 \qquad (22)$$

$$\sigma_{yy}(\omega) = \frac{i}{2\pi^2} \left( \frac{t + t'}{\sqrt{3t}} \int d\theta d\varepsilon \, \frac{1}{2\varepsilon + \omega + i\alpha} \right.$$
$$\left. \times \frac{(\varepsilon\cos\theta - \Delta)^2}{\sqrt{\varepsilon\cos\theta} \left( \varepsilon^2 + \Delta^2 - 2\Delta\varepsilon\cos\theta \right)} \right) \sigma_0 \qquad (23)$$

As $t' = 2t$, the energy gap parameter approaches zero. The real part of the longitudinal conductivities long x-direction yield $\sigma_{xx}'(\omega) \approx \sigma_0 \left[ \sqrt{3t/2\sqrt{\omega}} \right]/(6t\pi)$ and the y-direction $\sigma_{yy}'(\omega) \approx \sigma_0 \left[ 3t\sqrt{2/(3t)} \right]/(16\pi\sqrt{\omega})$, corresponding to those findings in Ref.[15].

## V. TRANSMITTANCE AND ABSORBANCE

In this section, we examine the interaction between electron in a hexagonal lattice under various conditions of $t'$ and an external electromagnetic wave of a specific frequency $\omega$. We solve Maxwell's equations for the polarized electromagnetic wave propagating along the $z$ axis and interacting with electrons in hexagonal lattice at $z=0$, as illustrated in Fig.5. The Maxwell's equations reads [34]

$$\frac{\partial^2 \mathbf{E}}{\partial z^2} = \frac{\omega^2}{c^2}\varepsilon\mathbf{E} - 4\pi i\frac{\omega}{c^2}\delta(\omega)\mathbf{j} \quad (24)$$

The discontinuity and continuity of the tangential electric field are expressed as

$$-iq\mathbf{E}_t + iq(\mathbf{E}_i - \mathbf{E}_r) = 4\pi i\frac{\omega}{c^2}\mathbf{j} \quad (25)$$

$$(E_x^x, 0) + (E_r^x, E_r^y) = (E_t^x, E_t^y) \quad (26)$$

By applying Ohm's Law $\mathbf{j} = \sigma\mathbf{E}$, the transmission and reflection amplitudes can be represented as follows [15]

$$\frac{E_t^{\mu}}{E_i^{\mu}} = \frac{1}{1 + \dfrac{2\pi}{c}\sigma_{\mu\mu}} \equiv t_{\mu} \quad (27)$$

$$\frac{E_r^{\mu}}{E_i^{\mu}} = -\frac{\dfrac{2\pi}{c}\sigma_{\mu\mu}}{1 + \dfrac{2\pi}{c}\sigma_{\mu\mu}} \equiv r_{\mu} \quad (28)$$

with the conditions of $E_i^x = |E_i|\cos\theta$ and $E_i^y = |E_i|\sin\theta$. The transmittance and reflectance read

$$T = |t_x|^2 + |t_y|^2, \quad R = |r_x|^2 + |r_y|^2 \quad (29)$$

Fig.6 illustrates the transmittance under various conditions of $t'$ and polarization angle $\theta$. When the polarization angle $\theta = 0^o$ and $t'/t < 1$, the transmittance decreases, but it enhances with rising $t'/t$ from the original condition of $t'=t$ in low-energy limit. At $t'/t > 2$, the material exhibits optical transparency as long as the energy of the electromagnetic wave remains below the energy gap. However, as the polarization angle increases to $\theta = 90^o$, the transmittance exhibits the opposite trend compared to its behavior at $\theta = 0$. Furthermore, the transmittance as a function of polarization angle is illustrated in Fig.7. When the light energy is fixed at $\omega/t = 0.1$, the transmittance reduces at the polarization angle $\theta = 90^o$ for $t'/t > 1$, but increases in the case of $t'/t < 1$. Since $\omega/t$ lies within the gap when $t'/t = 2.5$, the electron transitions from the valence band to the conduction band are prohibited, resulting in perfect transmission. Interestingly, the transmittance reduces to 86% when $t'/t = 0.5$ and $\omega/t = 1$, as illustrated in Fig.7b. The dip of transmittance was further analyzed for $t'/t = 2$, as discussed in Ref.[15]. The study revealed that a significant dip appears when reaches very low values of $\omega/t$. These findings suggest the use of strain to control the light transmission : $t'/t < 1$ for the visible regime and $t'/t = 2$ for the microwave and far-infrared regime, under the assumption of $t \sim 2.8$ eV.

The reflectance is negligible in comparison to the transmittance. Therefore, the absorbance can be defined as $A = 1 - R - T$. It is evidence that the absorbance is large when $t'/t = 2$ and the polarization angle $\theta = 90^o$ for low energy of $\omega/t \sim 0.1$, as illustrated in Fig8a [15]. On the other hand, significant absorbance is observed when $t'/t = 0.5$ within visible regime ($\omega/t \sim 1$) and polarization angle of $\theta = 0^o$, as shown in Fig.8b. For a polarization angle of $\theta = 90^o$, this phenomenon is observed when $t'/t = 1.5$.

Fig.9 presents a comparison of absorbance across different $t'/t$ at a constant light energy $\omega/t$, for polarization angle of $\theta = 0^o$ and $\theta = 90^o$. The results clearly demonstrate that a single absorbance peak occurs at $\theta = 0^o$, whereas multi-peaks are observed at $\theta = 90^o$ [35].

The analytical expression of transmittance can be investigated analytically using the optical conductivities in Eq.18-19. We introduce the fine-structure constant $\alpha = e^2/(\hbar c) \sim 1/137$ along with a new variable $\sigma_{\mu\nu} \equiv \sigma_{\mu\nu}/\sigma_0$ to facilitate the analysis. In low-energy limit, the transmittance when $t'/t < 2$ can be approximated as follows

$$T \approx 1 - \pi\alpha\left(\sigma_{xx}\cos^2\theta + \sigma_{yy}\sin^2\theta\right)$$

$$\approx 1 - \pi\alpha \left[ \left( \sqrt{\frac{4t^2}{3t'^2} - \frac{1}{3}} \right) \cos^2\theta + \left( \frac{3t'}{t\sqrt{12 - \frac{3t'^2}{t^2}}} \right) \sin^2\theta \right] \tag{30}$$

The reflectance can likewise be approximated as follows

$$R \approx \left( \frac{\pi\alpha}{2} \right)^2 \left( \sigma_{xx}^2 \cos^2\theta + \sigma_{yy}^2 \sin^2\theta \right) \tag{31}$$

The reflectance scales with $\alpha^2$, whereas the transmittance is directly proportional to $\alpha$. Consequently, the reflectance can be disregarded, as noted in the full-band investigation. Therefore, the absorbance in the low-energy regime can be approximated as follows

$$A \approx \frac{\pi\alpha \left( 2t^2 + t'^2 + 2(t^2 - t'^2)\cos 2\theta \right)}{tt'\sqrt{12 - \frac{3t'^2}{t^2}}} \tag{32}$$

The analytical expression of absorbance in Eq.32 within the low-energy limit $\omega/t \sim 0.5$ aligns with the findings presented in Fig.9, revealing absorbance values of $A \sim \pi\alpha\sqrt{4-x^2}/(\sqrt{3}x)$ for the polarization angle of $\theta = 0^o$ and $A \sim \sqrt{3}\pi\alpha x/\sqrt{4-x^2}$ for $\theta = 90^o$, where $x \equiv t'/t \leq 1.5$.

Unfortunately, the analytical expression for transmittance and absorbance when $t \geq 2$ are complicated, and the results are therefore not presented.

## VI. M-POINT-SADDLE-RESOLVED CONDUCTIVITY

The van Hove singularities associated with M-points exhibit significantly strong optical transport in both longitudinal conductivity components $\sigma_{xx}$ and $\sigma_{yy}$. There exist three inequivalent M-point saddles located within Brillouin zone, as illustrated in Fig.2c. The concept of saddle polarization — analogous to the valley polarization at K and K' — is introduced beyond the regime of linear dispersion around nodal points. This polarization can be actively controlled using linearly polarized light [36], considering the full band structure. Among the three inequivalent M-points when $t' = t$, two ($M_1$ and $M_2$) share identical conductivity characteristics, while the third ($M_3$) presents a distinct behavior, as shown in Fig.10a and 10b.

The excitation rate of electrons transitioning from the valence band to the conduction band at M-points, driven by the absorption of linearly polarized light with a specific polarization angle $\theta$, can be determined as follows [37, 38]

$$W_{v \to c} = \frac{2\pi}{h} |M|^2 \delta(2E_k - \omega) \tag{33}$$

where the term $M \equiv \langle \psi_+ | \nabla_\mathbf{k} H_0 | \psi_- \rangle \cdot \mathbf{A}$ represents the perturbation introduced by the linearly polarized light.

The M-point-saddle-resolved analysis reveals notable optical insulation along the x-axis near a particular $M_3$-point (black-dashed lines), as depicted in Fig.10a. This arises from the fact that the band structure (conduction band) at this $M_3$-point exhibits a local maximum along the x-axis, preventing electron transitions via x-polarized light. Further investigation using transition rates, as described by Eq.33, across different polarization angles of light highlights this phenomenon. Specifically, electron transitions induced by x-polarized light with an energy of $\omega = 2t$ are permitted only near two of the three M-points ($M_1$ and $M_2$), whereas the third M-point ($M_3$) remains inaccessible, as illustrated in Fig.10c. Additionally, a single M-point saddle exhibits a negative response when the polarization angle is set to $0^o$, $\pm 60^o$, $\pm 120^o$ [39]. This can be interpreted as the polarization angle of $\theta = 0^o$ suppressing electron transitions at $M_3$, $\theta = 60^o$ at $M_2$, and $\theta = 120^o$ at $M_1$.

The unexpected behavior of saddle polarization becomes particularly evident in the strain-modulated selection of M-point saddles. In this scenario, we analyzed the case of $t' = 0.5t$, which demonstrates a nearly perfect M-point saddle filtering effect at a light energy of $\omega = 3t$ under y-polarized light across the region of $M_3$ in M-point-saddle-resolved optical conductivity, as illustrated in Fig.11b.

A comparison of excitation rates between Fig.10c (without strain) and Fig.11c (with strain) when $\theta = 0^o$ reveals the suppression of optical transitions near $M_1$ and $M_2$ due to strain effects, as indicated by the color scale.

A detailed examination of these behaviors can be conducted along the line $K \rightarrow M \rightarrow K'$, as shown in Fig.12a. At $t'=t$ and the polarization angle of $\theta = 90^o$, the transition rate is dominant around the M-point while diminishing at the other locations. The dependence of the transition rate at each M-point on the polarization angle $\theta$ exhibits a periodic behavior, as depicted in Fig.12b. They exhibit the same amplitude in their periodic function but with a phase delay.

Notably, when $t'=0.5t$ and $\omega = 3t$, the transition rate at $M_3$ remains dominant, whereas at $M_1$ and $M_2$, it approaches zero, as illustrated in Fig.12c. The small transition rate magnitudes at $M_1$ and $M_2$ are further visualized in Fig.12d, confirming the presence of an almost perfect M-point saddle filtering effect. The analysis indicates that this filtering effect is most pronounced when the polarization angle deviates from x-polarized light, optimizing the transition rate under y-polarized light.

## VII. DISCUSSION AND CONCLUSIONS

The investigation of electronic and optical properties of an anisotropic hexagonal lattice are explored using the tight-binding Hamiltonian. Lattice deformation results in the omnidirectional transport characteristics. Compression occurs along the armchair orientation can be characterized by $t'/t > 1$. Conversely, when $t'/t < 1$, tension is directed along the armchair direction [9, 40].

In the case of $t'/t < 2$, the Fermi velocity around the nodal points along the x-axis reads $v_F^{(x)} \approx (at/2)\sqrt{12 - (3t'^2)/t^2}$ while along the y-axis becomes $v_F^{(y)} \approx 3at'/2$. On the other hand, it becomes a massive particle along the x-axis with its mass of $m^* \approx 2/(3a^2 t)$ [11], while maintaining massless particle with Fermi velocity $v_F^{(y)} \approx a(t + t')$ along the y-axis when $t'/t \geq 2$. Note that the energy gap $\Delta \equiv t' - 2t$ opens when $t'/t > 2$, which two nodal points merge each other.

The DOS can be manipulated using the strain parameter $t'/t$, as demonstrated in Fig.3. In the low-energy limit of $t'/t < 2$, the DOS exhibits a linear dependence on the energy level and reaches its minimum under the condition of $t' = \sqrt{2}t$. When $t'/t \geq 2$, the DOS relies on the square root of the energy level with the threshold energy $\Delta \equiv t' - 2t$ [11, 29].

The longitudinal conductivities associated with each spin and valley projection across different directions yield a scalable factor of the universal conductivity $\sigma_0$, given as $\sqrt{4 - x^2}/(\sqrt{3}x)$ for $\sigma_{xx}'(\omega)$ and $(\sqrt{3}x)/\sqrt{4 - x^2}$ for $\sigma_{yy}'(\omega)$ when $x \equiv t'/t < 2$. As $t'/t$ increases, lattice deformation boosts conductivity along the y-axis while reducing it along the x-axis. Under the condition of $t'/t \geq 2$, significant conductivity along y-axis is observed, being proportional to $\sqrt{\omega}$. Conversely, the small conductivity along the x-axis relies on $1/\sqrt{\omega}$, with a threshold energy of $2|\Delta|$. This result aligns with the findings on the product of $\sqrt{\sigma_{xx}(\omega)\sigma_{yy}(\omega)} \sim \sigma_0$, reported in Ref.[12, 15].

The transparency of polarized light is among a significant effect of omnidirectional transport. The deep of the transmittance dip, as a function of the light energy, varies with the rotation of the plane of polarization. These dips arise from the $M$-points, where the associated energies are positioned at $\omega = 2|t'|$ and $\omega = 2|2t - t'|$. Within low energy of light $\omega/t \sim 0.1$ (infrared regime), light is almost entirely transmitted across all cases of $t'/t$ except when $t'/t = 2$ and the polarization angle aligns with the armchair direction ($\theta = 90^o$) [15]. Conversely, a layered material exhibits translucency in visible regime ($\omega/t \sim 1$) when $t'/t \sim 1.5$ with the polarization angle aligned along the armchair direction ($\theta = 90^o$), and $t'/t \sim 0.5$ with the polarization angle aligned along the zigzag direction ($\theta = 0^o$). Notably, this discovery reveals greater fluctuations in transmittance compared to those reported in Ref.[14, 40, 41]. The findings suggest that strain on hexagonal lattice paves the way for the visible light-harvesting materials, where the large absorption of light energy can be tuned by adjusting the strain parameter.

The tunable hopping parameter $t'/t$ can be compared with the model of lattice deformation, described in Ref. [9]. The calculation is performed by evaluating the energy level at $M$-point, corresponding to the van Hove singularity. For compression along the armchair direction, the strength of $t'/t$ is equivalent to the tensile strain $\varepsilon$ by $(t'/t = 1.0, \varepsilon = 0)$, $(t'/t = 1.5, \varepsilon \approx 0.1)$, $(t'/t = 2.0, \varepsilon \approx 0.23)$, and $(t'/t = 2.5, \varepsilon \approx 0.39)$. Conversely, tension along the armchair direction reaches saturation around ($t'/t \approx 0.69, \varepsilon \approx 0.25$), beyond which the energy level at $M$ -point cannot be further increased within the framework of the model proposed in Ref. [9].

In pristine hexagonal system ( $t' = t$ ), the M-point-saddle-resolved optical response reveals a notable phenomenon: two of the M-points exhibit identical conductivity, while the third displays a distinct behavior, controlled using linearly polarized light. When the polarization angle aligns with the edge of hexagonal Brillouin zone, the optical transition at M-point located at that edge is prohibited, whereas the other two M-point saddles remain allowed.

In the presence of strain on hexagonal lattice $t' = 0.5t$ , a nearly perfect M-point saddle filtering effect is observed at the $M_3$ -point, while other two M-points are suppressed. This phenomenon gives rise to a strong transition rate at a polarization angle of $\theta = 90^o$ , corresponding to the armchair direction of the lattice. Conversely, when the polarization angle of $\theta = 0^o$ aligns with zigzag direction of lattice, the M-point saddle filtering effects is completely eliminated.

This strain model, applied to hexagonal lattice, could potentially extend to other two-dimensional materials such as phosphorene, with the model of $t'/t \geq 2$ [25, 42-45]. Research reveals that phosphorene absorbs light primarily in the infrared spectrum [16, 17]. Notably, the tunable band gap in phosphorene, achieved through doping with transition metal dichalcogenides (TMDs) [24] and TMDs/phosphorene heterostructures [46], improves its optical response, from infrared to visible regime. Additionally, the $d-$ orbital of transition metal atoms introduce substantial spin-spin interactions in phosphorene, enabling adjustable energy gaps [19, 47, 48]. Another example case of $t'/t < 1$ could apply for borophene oxide [26, 27, 49]. The findings from the first-principles investigation suggest that variations in the concentration of oxygen adatoms can alter the ratio of $t'/t$ [50], leading to a prediction of significant light absorption in the near-infrared region.

Mechanical strain can be directly utilized to enhance device performance across a broad range of light energies. The charge excitation configuration at M-point saddles, stimulated by linearly polarized light, offers an alternative method for encoding the quantum states, distinct from the valley polarization, which can be observed through the current flow [39]. These findings highlight the potential for controlling electronic and optical properties[48, 51, 52], contributing to the design of optoelectronic and straintronic devices.

In summary, we have presented a comprehensive theoretical study of strain-tuned optical properties in a two-dimensional hexagonal lattice. The main findings are:

1. Uniaxial strain controls the merging of Dirac cones, induces a tunable energy gap, and modifies the density of states.
2. Longitudinal optical conductivities become strongly anisotropic, leading to polarization-dependent transmittance and absorbance.
3. A novel M-point saddle filtering effect and saddle polarization emerge, enabling selective control of optical transitions beyond conventional valley polarization.
4. The predicted effects occur within strain and photon-energy regimes achievable in current 2D materials, including black phosphorus and borophene oxide.

These results open a pathway to strain-programmable optoelectronic devices, such as polarization-selective photodetectors, tunable absorbers, and ultrathin optical filters. Future work should focus on first-principles calculations and experimental verification of M-point saddle filtering in specific materials, as well as the integration of this effect into device architectures for real-world applications.

## ACKNOWLEDGMENTS

Phusit Nualpijit acknowledges the short-term collaboration on the topic of strain in graphene with Prof. Klaus Ziegler and Prof. Andreas Sinner during the time as a master's and Ph.D. student and also appreciates Marion Amling's suggestion regarding living in Augsburg, Germany. This work was supported by Kasetsart university research and development institute (KURDI) FF(KU-SRIU) 7.67.

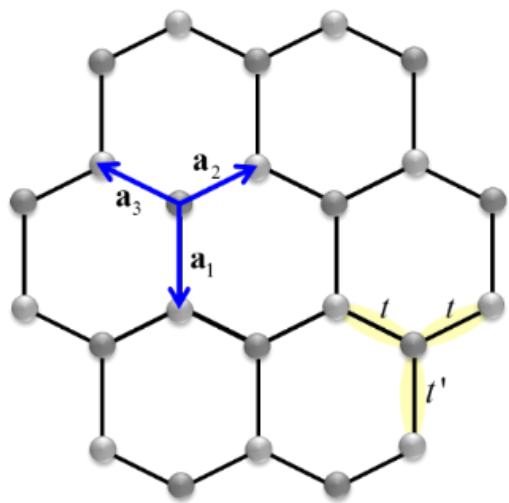

Fig.1: The hexagonal lattice with the nearest-neighbor vectors $\mathbf{a}_i$ and hopping parameters $t_i$.

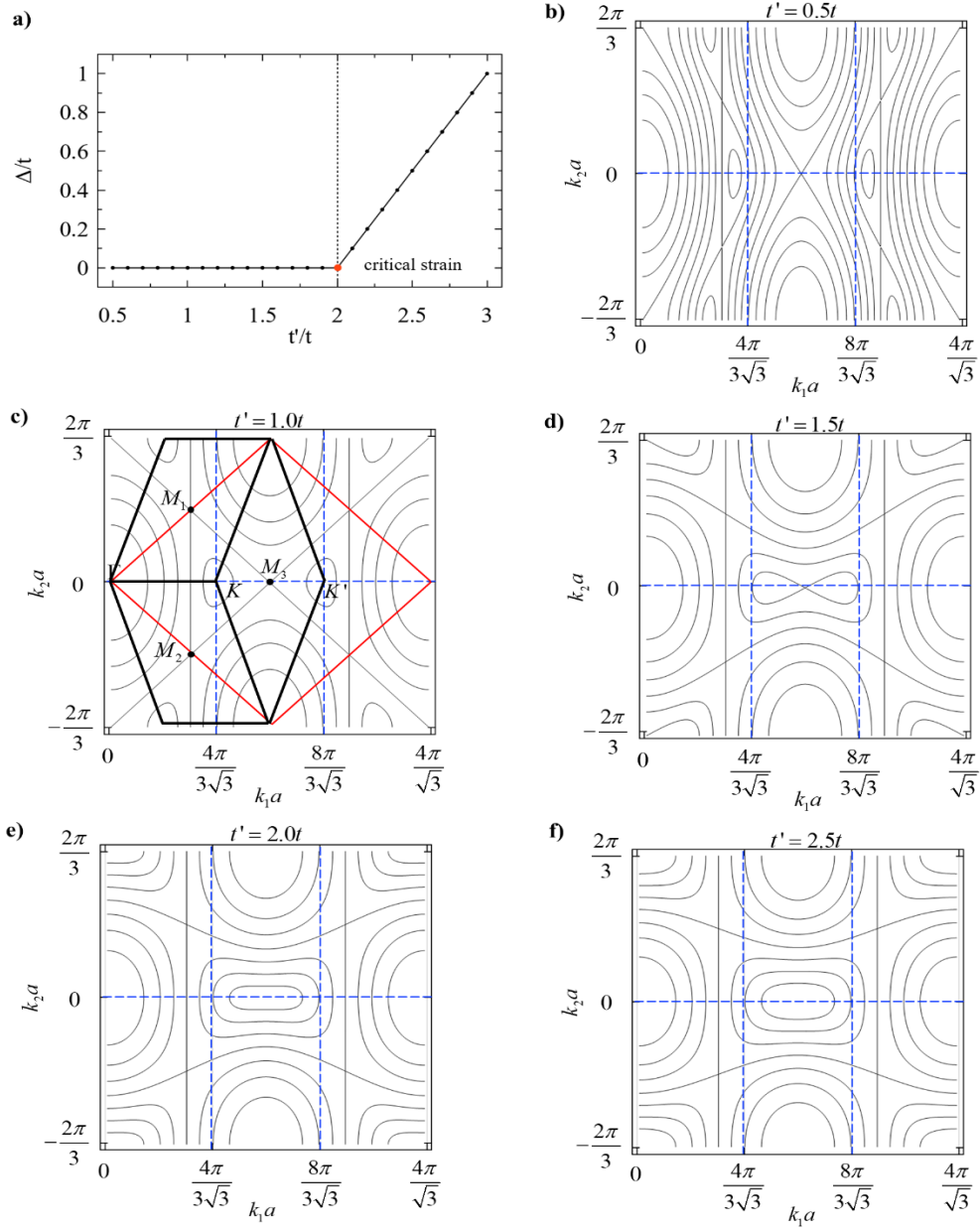

Fig.2: (a) the energy gap parameter $\Delta$, as a function of $t'$. (b)-(e) Contour plots of the energy dispersion relation for $t'=0.5t$, $t'=1.0t$, $t'=1.5t$, $t'=2.0t$, and $t'=2.5t$, respectively. The area enclosed by the red diamond in (b) represents the area of Brillouin zone, bounded by four $\Gamma$ points. The black diamond marks to M-point resolved within Brillouin zone.

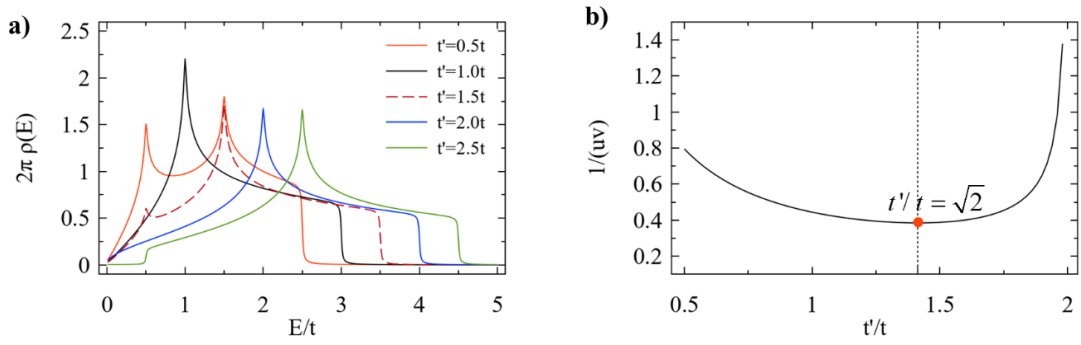

Fig.3: (a) The density of states $\rho(E)$ as a function of $E/t$ across various hopping parameter $t'$. (b) scale function $1/|uv|$ as a function of $t'/t$.

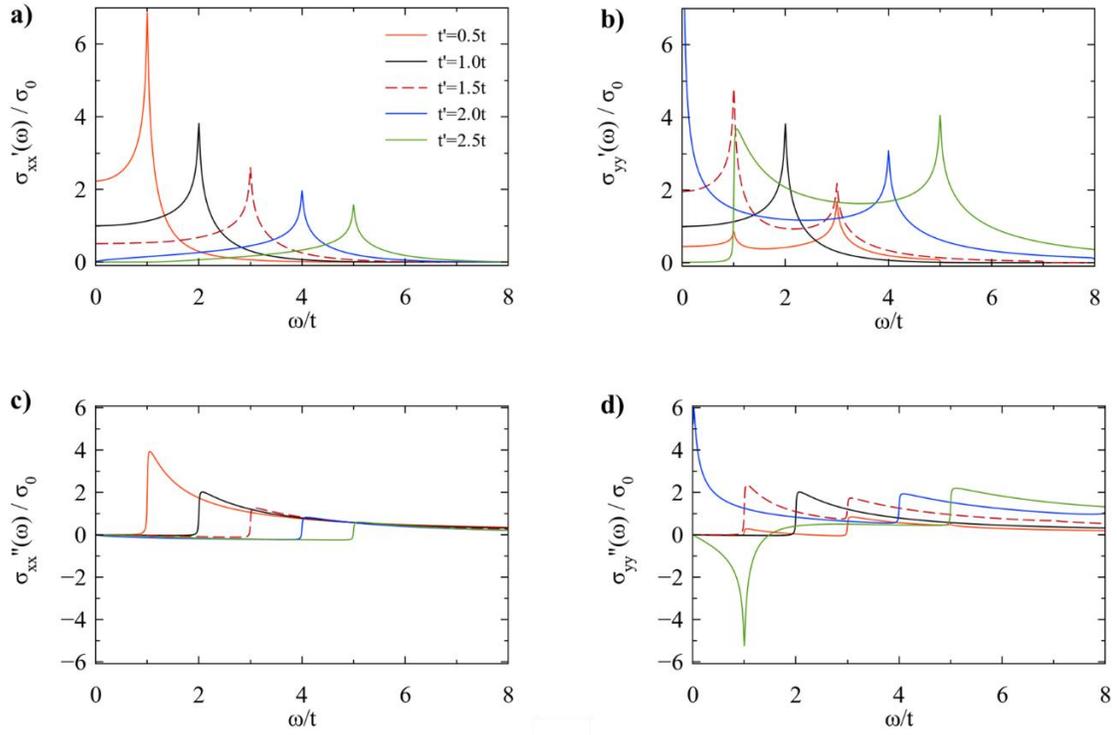

Fig.4: The longitudinal optical conductivities for the real part (a) $\sigma_{xx}{}'(\omega)$, (b) $\sigma_{yy}{}'(\omega)$ and the imaginary part (c) $\sigma_{xx}{}''(\omega)$, (d) $\sigma_{yy}{}''(\omega)$ as a function of light energy $\omega$.

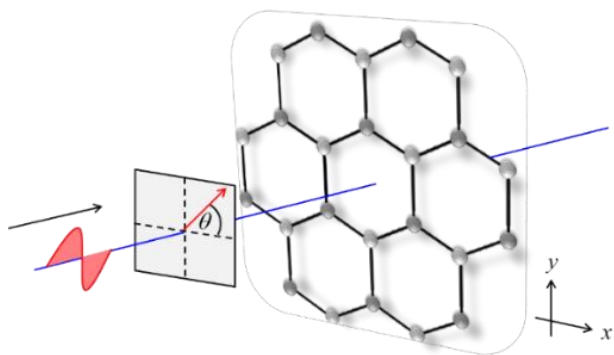

Fig.5: The propagation of linearly polarized light along the z axis with a polarization angle of $\theta$ relative to the x axis. When $\theta = 0^o$, the polarization angle aligns with the zigzag direction of hexagonal lattice.

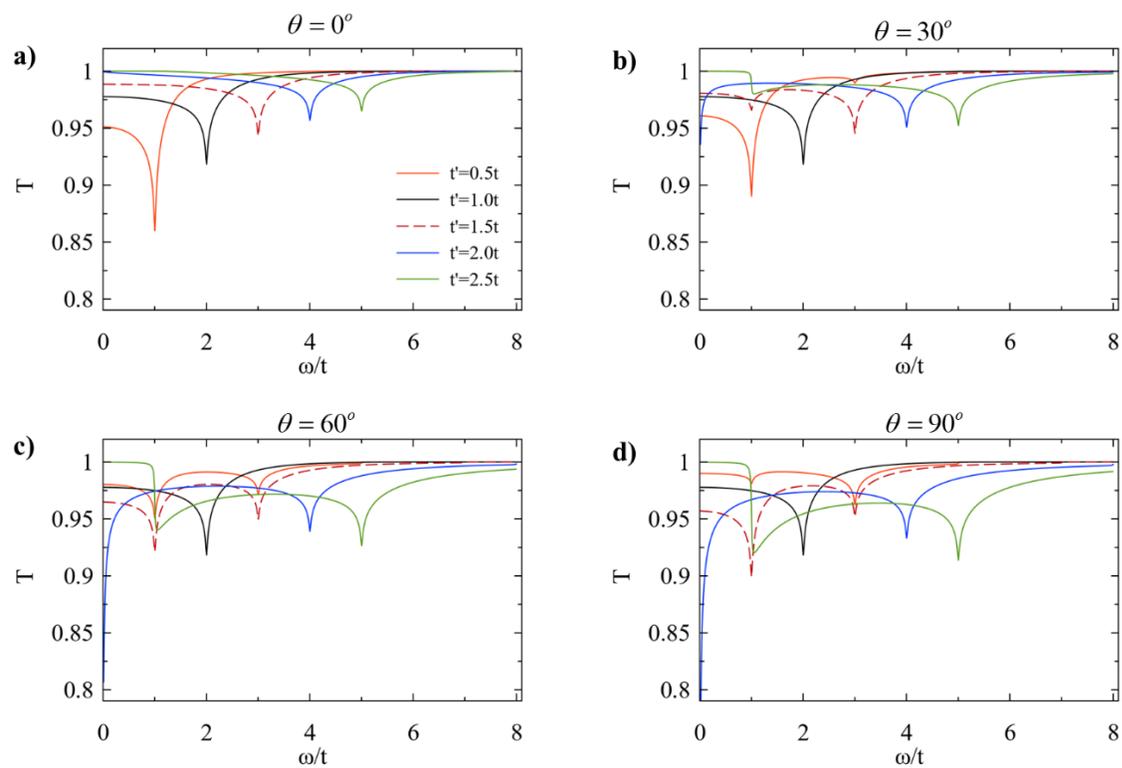

Fig.6: The transmittance across various conditions of $t'$ as a function of light energy when the polarization angle (a) $0^0$, (b) $30^0$, (c) $60^0$, and (d) $90^0$.

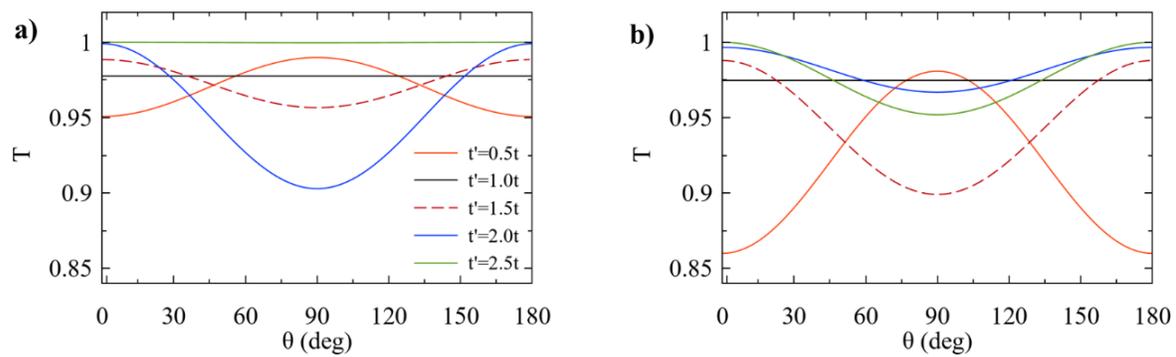

Fig.7: The transmittance as a function of polarization angle across various $t'/t$ for (a) $\omega/t = 0.1$, and (b) $\omega/t = 1$.

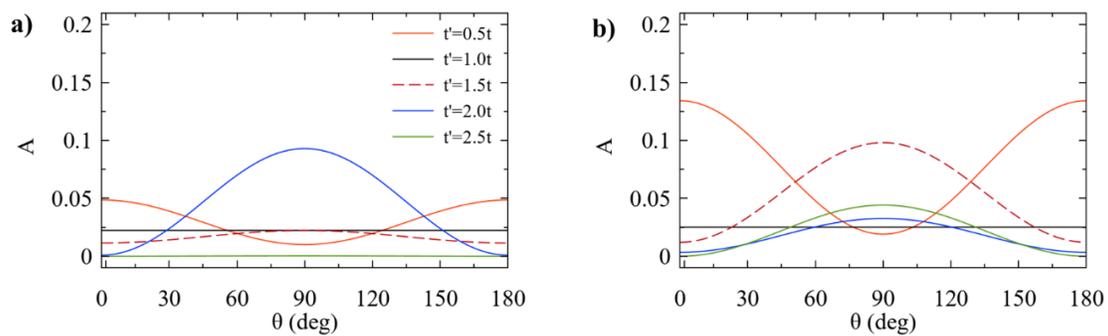

Fig.8: The absorbance as a function of the polarization angle of incident light across various $t'/t$ for (a) $\omega/t = 0.1$, and (b) $\omega/t = 1$. In low energy regime ($\omega/t = 0.1$), large absorbance occurs when $t'/t = 2$, whereas in higher energy regime ($\omega/t = 1$), a large absorbance occurs when $t'/t = 0.5, 1.5$.

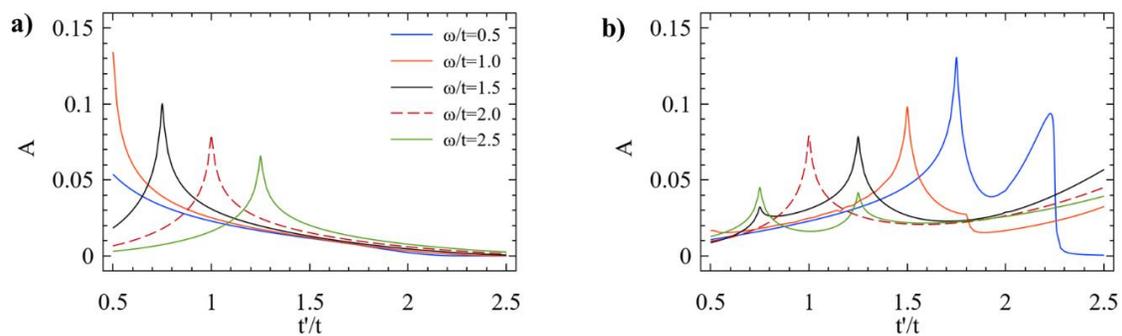

Fig.9: The absorbance as a function of strain parameter $t'/t$ across various $\omega/t$ at the polarization angle of (a) $\theta = 0^o$ and (b) $\theta = 90^o$.

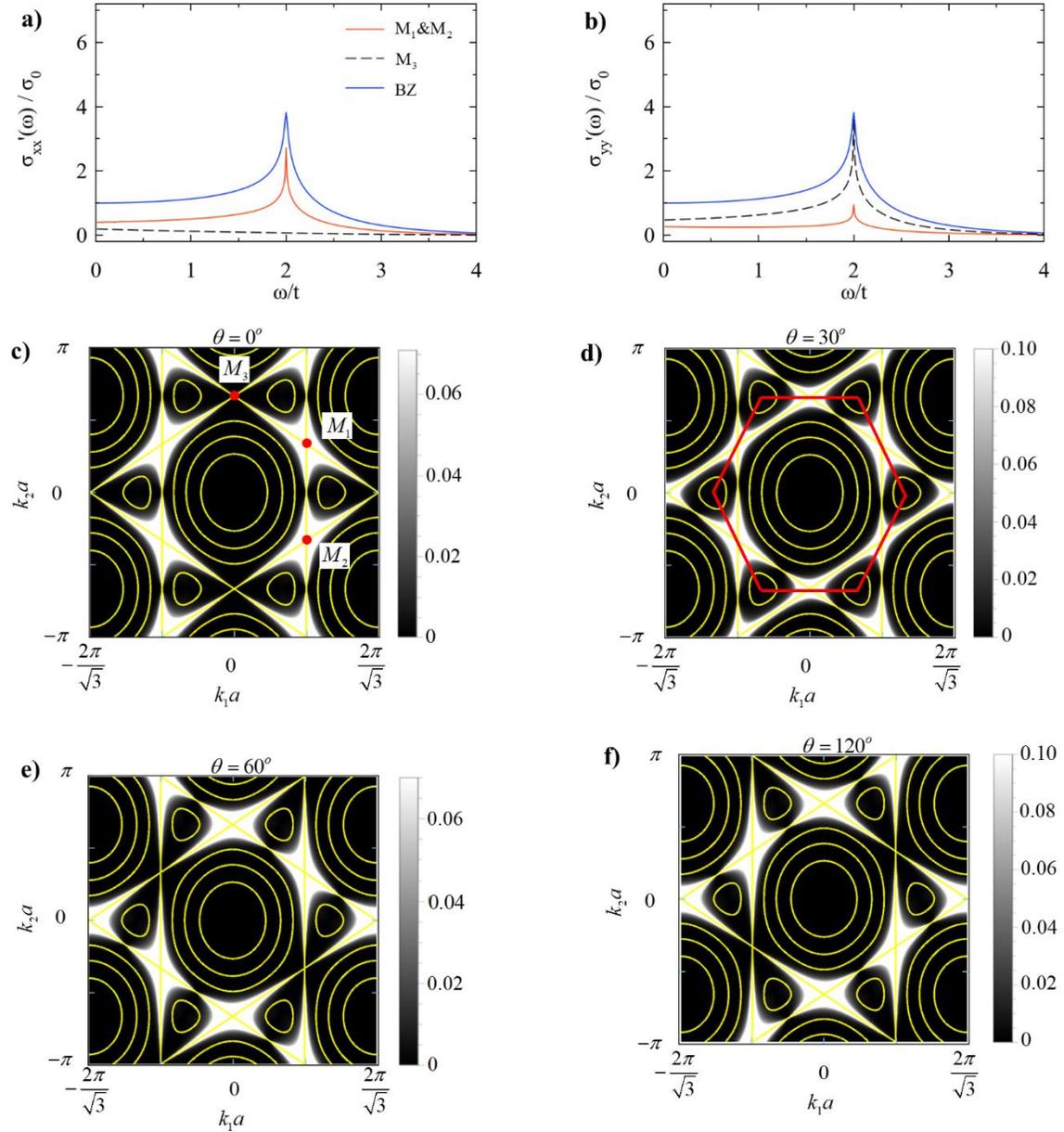

Fig.10: The M-point-saddle-resolved conductivities (a) $\sigma_{xx}'$, (b) $\sigma_{yy}'$ when $t'=t$. The transition rate of electron from valence band to conduction band (represented by color scale) at the light energy of $\omega=2t$ when the polarization angle is (c) $\theta=0^o$ (negative response at $M_3$), (d) $\theta=30^o$, (e) $\theta=60^o$ (negative response at $M_2$), and (f) $\theta=120^o$ (negative response at $M_1$). The negative response of electrons transitioning occurs when the polarization angle aligns along the edge of hexagonal Brillouin zone (red line in (d)). The yellow contour lines refer to the energy level in Brillouin zone as same as in Fig.2c. Note that the transition rate at M-point is very large compared to the color scale.

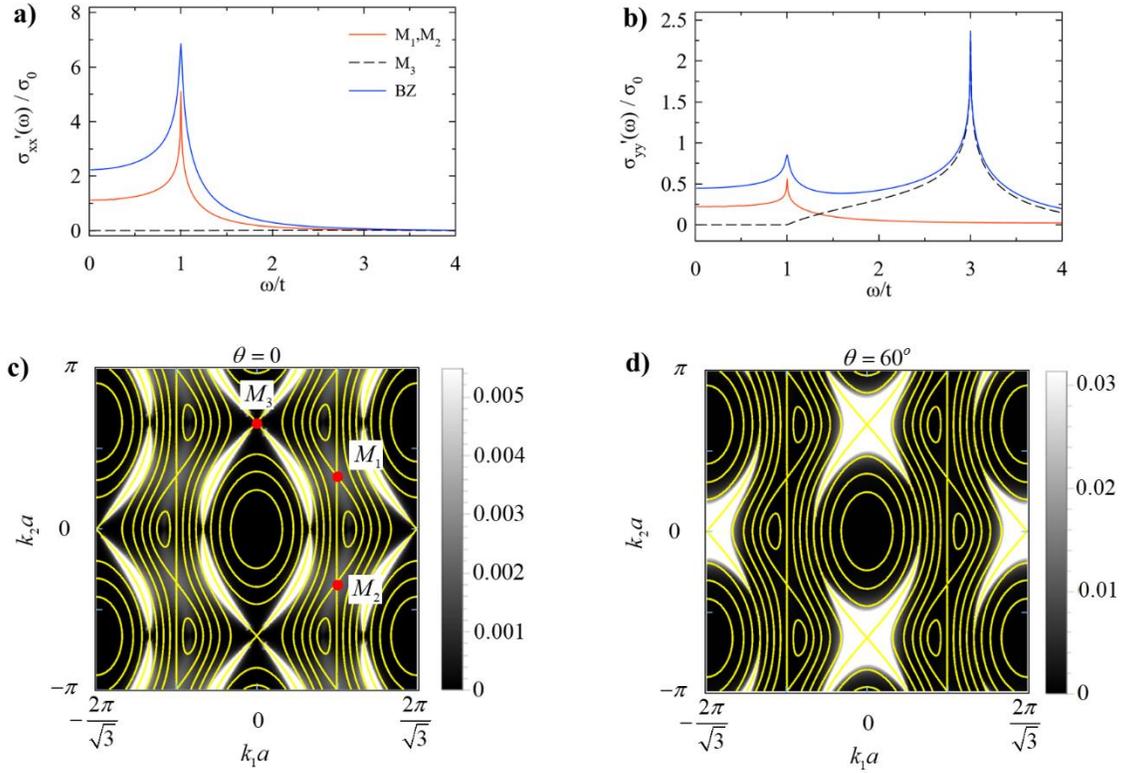

Fig.11: The M-point-saddle-resolved conductivities in (a) $\sigma_{xx}'$, (b) $\sigma_{yy}'$ when $t'=0.5t$. The disappearance of $\sigma_{xx}'$ in (a) at the $M_3$ indicates a negative response to x-polarized light. The transition rate of electrons from valence band to conduction band (represented by color scale) at the light energy $\omega=3t$ when the polarization angle is (c) $\theta=0^o$, and (d) $\theta=60^o$. The black area around $M_1$ and $M_2$ in (d) refers to the negative response of M-point saddles to the linearly polarized light with polarized angle of $\theta$.

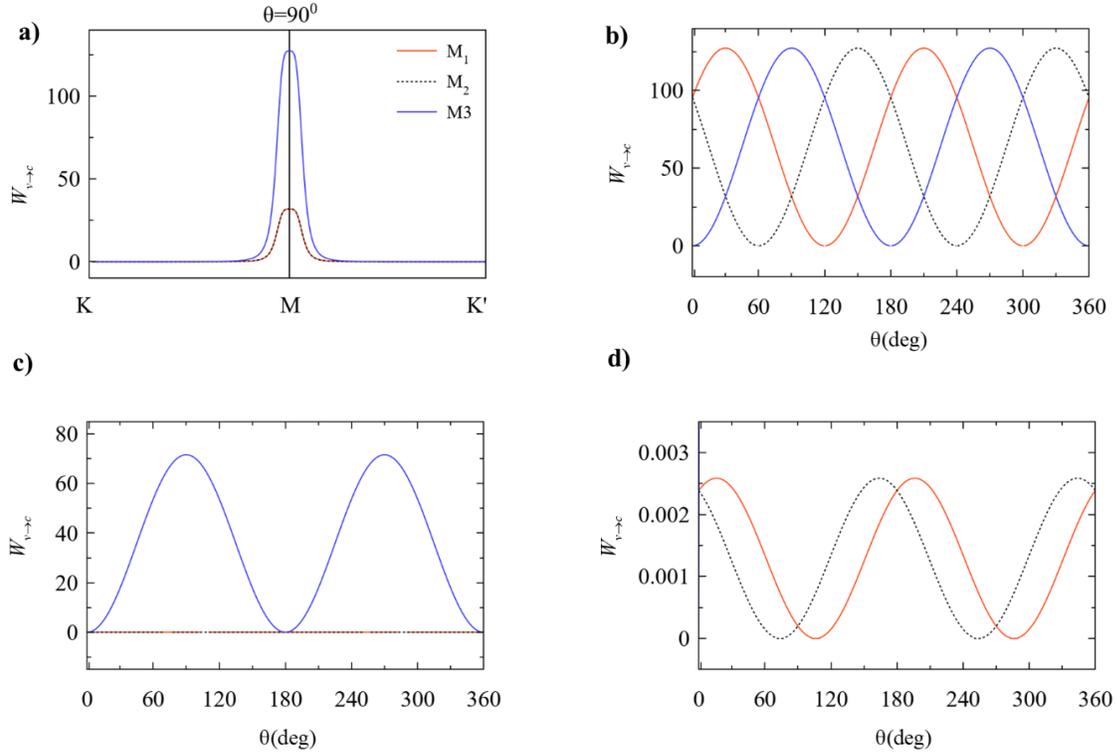

Fig.12 : (a) the M-point-saddle-resolved transition rate of electrons $W_{v \to \infty}$ along $K \to M \to K'$, stimulated by polarized light at an angle of $\theta = 90^o$. A high concentration of transitioning electrons is observed locally around M-points. The transition rate at M-point when (b) $t' = t$. (c), and (d) $t' = 0.5t$. The rate at $M_1$ and $M_2$ are significantly lower compared to $M_3$, so the plot is separated individually in (d). The maxima of (a) correspond to the rate in (b) when $\theta = 90^o$.